\documentclass[conference]{IEEEtran}

\IEEEoverridecommandlockouts                              

\usepackage{multirow}
\usepackage[table,xcdraw]{xcolor}
\usepackage{graphicx}
\usepackage{subcaption}

\title{\LARGE \bf
Musinger: Communication of Music over a Distance with Wearable Haptic Display and Touch Sensitive Surface
}

\author{Miguel Altamirano Cabrera, Muhammad Haris Khan, Ali Alabbas, \\ Luis Moreno, Issatay Tokmurziyev, and Dzmitry Tsetserukou
\thanks{All the authors are with the Intelligent Space Robotics Laboratory, Skolkovo Institute of Science and Technology (Skoltech),
         Bolshoy Boulevard, 30, p.1, Moscow 121205, Russia
        {\tt\small \{haris.khan, m.altamirano, ali.alabbas,  luis.moreno, issatay.tokmurziyev  d.tsetserukou}@skoltech.ru}}%

\begin{document}

\maketitle

\begin{abstract}

This study explores the integration of auditory and tactile experiences in musical haptics, focusing on enhancing sensory dimensions of music through touch. Addressing the gap in translating auditory signals to meaningful tactile feedback, our research introduces a novel method involving a touch-sensitive recorder and a wearable haptic display that captures musical interactions via force sensors and converts these into tactile sensations. Previous studies have shown the potential of haptic feedback to enhance musical expressivity, yet challenges remain in conveying complex musical nuances. Our method aims to expand music accessibility for individuals with hearing impairments and deepen digital musical interactions. Experimental results reveal high accuracy ($98\%$ without noise, $93\%$ with white noise) in melody recognition through tactile feedback, demonstrating effective transmission and perception of musical information. The findings highlight the potential of haptic technology to bridge sensory gaps, offering significant implications for music therapy, education, and remote musical collaboration, advancing the field of musical haptics and multi-sensory technology applications.

\end{abstract}


\section{INTRODUCTION}

Musical haptics has seen significant advancements as researchers explore the intersection of auditory and tactile experiences in music performance and perception. Musical haptics is an interdisciplinary area that integrates principles from haptic engineering, human-computer interaction (HCI), applied psychology, and musical acoustics to enhance the sensory dimensions of music through touch and proprioception \cite{Papetti2018}. This burgeoning field has led to the development of innovative devices and systems that aim to provide meaningful haptic feedback to musicians and audiences, thereby enriching the overall musical experience.

Previous research in this domain has demonstrated the potential of haptic feedback to improve the expressivity and control of digital musical instruments (DMIs). For example, the Laptop Orchestra of Louisiana has employed force-feedback instruments to provide performers with precise, physically intuitive, and reconfigurable controls. These instruments enable musicians to interact with digital music interfaces in ways that closely mirror the tactile engagement of traditional acoustic instruments, thus enhancing performance accuracy and expressiveness.

\begin{figure}[htp!]
  \centering
  \includegraphics[width=0.48\textwidth]{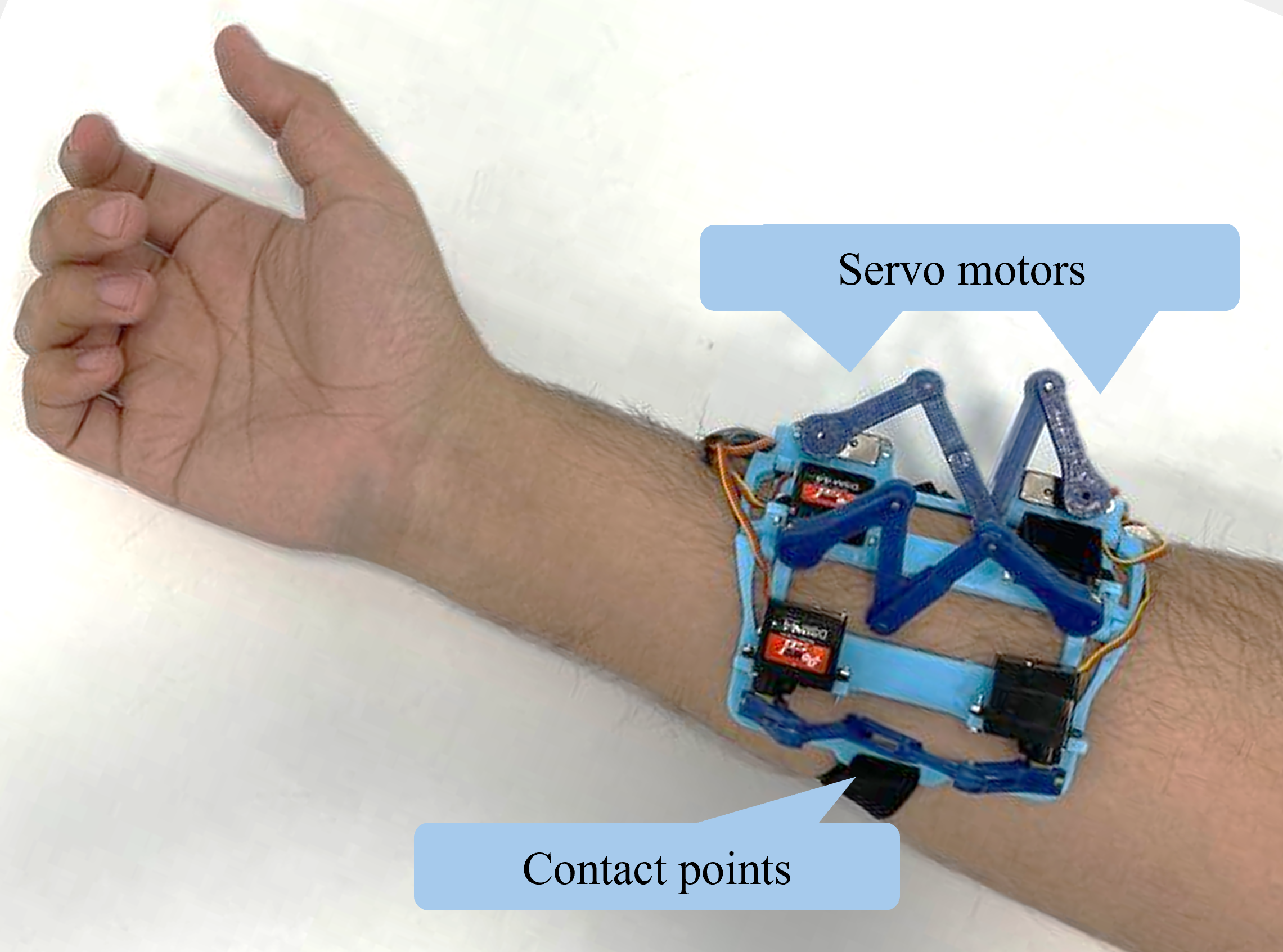}
  \caption{Wearable haptic display generating rhythm sensation on the user's forearm. The independent contact-points are generated by three inverse five-bar linkage mechanisms. }
  \label{fig:device}
\end{figure}

The integration of force feedback into DMIs has also been shown to reduce reaction times and enable more precise control of musical gestures, further blurring the lines between physical and digital music-making \cite{Berdahl2018}. Altukhaim et al. \cite{Altukhaim2024} demonstrated that synchronized haptic feedback significantly strengthens the sense of ownership over virtual body parts. Similarly, converting music into haptic feedback could deepen listeners' connections to music, allowing them to feel the music as well as hear it. This approach could revolutionize how we experience music, making it more immersive and accessible, particularly for those with hearing impairments. 

Tivadar et al. \cite{Tivadar2022} explore the effectiveness of digital haptic technologies in enhancing visual search performance in dual-task settings. Their study demonstrates that haptic-only feedback on touchscreen tablets can significantly reduce reaction times in visual tasks without compromising accuracy. This finding is particularly relevant for applications where visual attention is critical, such as driving, by suggesting that haptics can alleviate visual overload. This underscores the potential of haptic technology to improve user performance and safety in multitasking environments, paving the way for its broader application in interactive systems, including music-to-haptics conversion where tactile feedback could similarly enhance user engagement and experience.

The study conducted by Turchet et al. \cite{turchet2021touching} effectively addresses several critical challenges in the domain of musical haptics, particularly emphasizing the necessity for compact, unobtrusive, and efficient sensors and actuators capable of delivering high-fidelity haptic feedback. By concentrating on the translation of auditory signals into tactile sensations, the research provides valuable insights into the ways haptic cues can shape the perception of music, especially in environments where auditory feedback is either absent or restricted. 

Building on these foundational studies, the present research seeks to advance the field of musical haptics by exploring the conversion of auditory musical signals into haptic feedback. Specifically, this study investigates a novel method where one individual listens to music through headphones while simultaneously tapping on force sensors. Simultaneously signals from these sensors are then transmitted to a wearable robotic device worn by another individual, who does not have access to the auditory component of the music. This system aims to translate the auditory experience of music into a tactile one, thereby enabling the second individual to ``feel" the music through haptic feedback alone. 

The objective of this paper is to present a system that converts auditory musical signals into haptic feedback. This paper is structured as follows: Section II reviews prior research, highlighting advancements and challenges in force-feedback instruments and haptic wearables. Section III elaborates on the system architecture, detailing the design and technical setup of the new haptic feedback system, including force sensors and wearable robotics. The experimental evaluation is covered in Section IV, where the methodology, data collection, and analysis assess the system's effectiveness. Section V discusses potential applications with a focus on its use in music communication, performance, and accessibility, while also illustrating the system’s broader impact across various domains. Finally, Section IV concludes the paper by summarizing the findings and contributions to the field and proposing directions for future research.

\section{RELATED WORK}
\subsection{Haptic Interfaces}

Haptic feedback and wearable devices have significantly enhanced interactive technologies, especially in fields requiring precise human-robot interactions (HRI). One of such a system is a stylus-controlled musical interface that integrates resistive force feedback to enrich user interactions with digital musical instruments by simulating different surface textures \cite{muller2010reflective}. This approach not only augments the tactile experience but also improves the expressive capabilities of musicians in live performances.

Building on the integration of haptic feedback in musical applications, a modular device offering rotary force feedback for digital musical instruments was presented \cite{kirkegaard2020torquetuner}. This innovation allows musicians to experience varying haptic effects that mimic traditional musical interactions, thereby deepening the engagement with digital instruments through physical feedback.

Meanwhile, another project focused on improving VR immersion by developing RecyGlide, a forearm-worn haptic display that aids in spatial awareness without interfering with hand-tracking systems \cite{heredia2019recyglide}. Their design includes inverted five-bar linkages for tactile feedback, optimizing user safety and interaction clarity during intensive VR applications.

TouchVR is a novel wearable haptic interface designed to enhance VR experiences by providing multimodal tactile feedback to the palm and fingertips \cite{10.1145/3355355.3361896}. The device uses an inverted Delta mechanism to generate 3D force vectors, simulating sensations like pressure and texture, while vibrotactile motors on the fingertips add realism. Integrated with HTC Vive Pro and Leap Motion, TouchVR significantly improves immersion in VR through applications like BallFeel and RoboX. The research highlights its potential in fields such as teleoperation and VR training, with future plans to incorporate thermal feedback.

 On another spectrum HaptiHug is an innovative haptic display designed to simulate the sensation of a hug over a distance, thereby enriching social interaction and emotional involvement in online communication \cite{10.1007/978-3-642-14064-8_49}. The system integrates the 3D virtual world Second Life, where it automatically recognizes and responds to "hug" cues from text messages by triggering a realistic hug sensation through the HaptiHug device. The device uses soft materials and rotating motors to generate pressure on the user’s chest and back, closely mimicking the experience of a human hug. User studies revealed that the HaptiHug successfully increased the immersion and emotional engagement of participants, making it a promising tool for enhancing the realism and emotional depth of online interactions.

For fingetip feedback, LinkTouch can be observed, a wearable haptic device designed to deliver high-fidelity two-degree-of-freedom (2-DoF) force feedback directly at the fingerpad \cite{6775473}. The device features an inverted five-bar linkage mechanism that enables precise control over the application of normal and tangential forces at any point on the fingertip. This innovative design facilitates a realistic sensation of making and breaking contact with virtual objects, significantly enhancing the user’s tactile experience in virtual environments. 

Considering more the user's state, iFeel IM is a system that enhances emotional communication in online interactions through haptic feedback corresponding to detected emotions \cite{5349516}. While their work demonstrates the potential of affective haptics to deepen emotional experiences, Musinger applies these principles to the domain of music. By translating auditory signals into tactile sensations, Musinger not only enhances the expressivity of music but also broadens its accessibility, offering a novel application of haptic feedback.

Lastly, the MoveTouch system can be analyzed, which combines motion capture and vibrotactile feedback to ensure safety in HRI \cite{alabbas2024movetouch}. The device, fitted on the user's wrist, features an adjustable fiducial marker visible to the motion capture system and uses vibration patterns to guide the user's movements, thereby preventing potential collisions in a shared space with robots.

\subsection{Haptics in Music}
Recent explorations at the intersection of music and haptic feedback highlight innovative approaches to enhancing musical experiences through tactile sensations. Papetti and Saitis \cite{Papetti2018} highlight the need for haptic feedback in digital musical instruments (DMIs) to enhance performance control and expressivity, addressing a gap where most DMIs fall short. The present work directly responds to this by translating auditory signals into multi-contact haptic feedback, improving both expressivity and accessibility in music. By advancing haptic feedback technology, Musinger builds on the challenges identified by Papetti and Saitis, offering a more immersive musical experience. Turchet et al. \cite{turchet2021touching} introduce Musical Haptic Wearables for Audiences (MHWAs), which enhance live music performances by providing audiences with vibrotactile feedback synchronized to the music. While their work demonstrates the potential for increased engagement and enjoyment, it also highlights challenges like the need for personalized experiences and precise synchronization. Musinger builds on these concepts by translating auditory signals into customizable haptic feedback, aiming to offer a more accessible and personalized musical experience beyond live performance contexts.

Großhauser and Hermann \cite{grosshauser2009augmented} present a wearable multi-modal sensor system designed for real-time audio-haptic feedback, primarily aimed at enhancing motion and posture sensing. Their work demonstrates the potential of integrating auditory and tactile feedback to improve user awareness and performance across various activities. Building on this, Musinger applies these principles specifically to music, where it translates auditory signals into haptic feedback. This approach broadens the application of haptic feedback by focusing on the translation of musical qualities into tactile sensations, thus enhancing the user experience in a novel context. In the domain of multisensory music experiences, Frid and Lindetorp \cite{frid2020haptic} investigate the integration of whole-body vibrations in musical installations. Their research indicates that preparatory workshops that familiarize composers with haptic technologies can significantly enhance the effectiveness of these multisensory experiences, encouraging innovative compositional practices that incorporate both auditory and tactile elements.

Armitage and Ng \cite{armitage2015configuring} explore the practical applications of designing haptic interfaces tailored for musical performance, highlighting the importance of synchronization between audio and haptic outputs. Their work presents a quantitative analysis of motor types, optimizing haptic feedback to improve the musical experience without distracting from the auditory elements. Lastly, Mazzoni and Bryan-Kinns \cite{mazzoni2016mood} introduce the 'Mood Glove', a haptic wearable that enhances film music perception by synchronizing vibrotactile feedback with audio cues. This technology demonstrates the potential of haptic feedback to enrich the emotional depth of film viewing, particularly for hearing-impaired audiences, by intensifying the perceived moods conveyed by the soundtrack.

These studies highlight the wide-ranging possibilities of using haptic feedback in music and performance arts, opening up exciting opportunities to enhance auditory experiences through touch. While earlier research focused on vibrotactile feedback, our study takes a different approach by utilizing multi-contact tactile sensations. We use a device that generates three independent contact points, allowing for more complex force vector sensations.

\section{SYSTEM ARCHITECTURE}
\begin{figure}[htp!]
  \centering
  \includegraphics[width=0.45\textwidth]{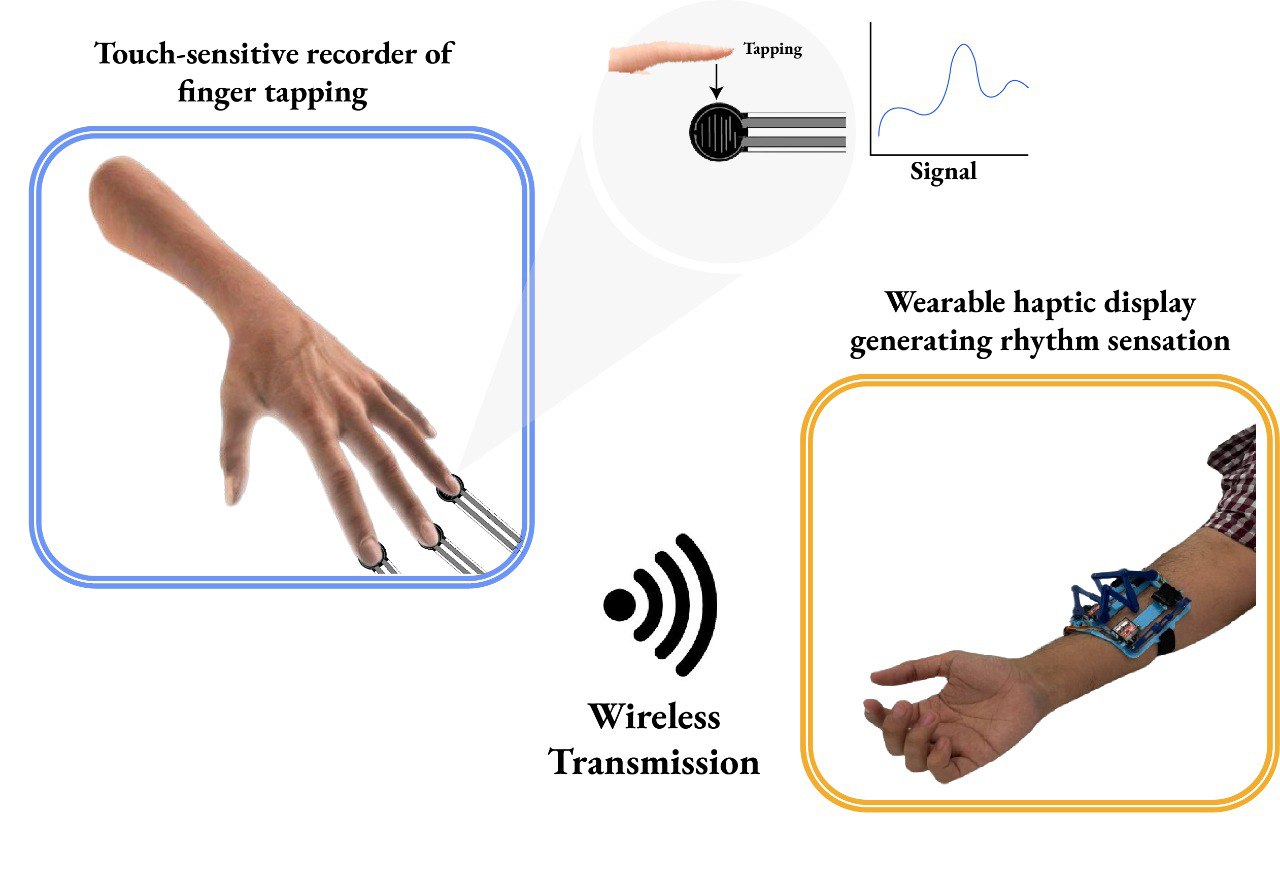}
  \caption{Musinger architecture.}
  \label{fig:arch}
\end{figure}

The proposed system consists of two main components. The first component is a touch-sensitive recorder that enables users to interact with music through tapping. It records these interactions using an array of force sensors, which are then transmitted to the second component: a haptic display. The haptic display is a wearable robotic device worn by a second user. The second user perceives the tactile sensations transmitted from the touch-sensitive recorder in real-time through tactile patterns rendered on their forearm. The system architecture is illustrated in Fig. \ref{fig:arch}. The second user is located in a remote environment, unable to see or hear the tapping sounds made by the first user.

\subsection{Touch-sensitive Recorder}
The touch-sensitive recorder comprises three single-point touch paths. Each contact point utilizes a Force Sensitive Resistor (FSR) SP200-10 from Peratech. The ergonomic placement of the three FSRs enables intuitive interaction using the index, middle, and ring fingers. As the user listens to music, they can tap the FSRs in sync with the music. These electrode signals are then captured by an ESP32 microcontroller, which records the data and subsequently transmits it to the wearable haptic display component.


\subsection{Wearable Haptic Display}

The wearable haptic display aims to create the sense of touch at three different points on the user's forearm. The device design is based on the inverted five-bar linkage mechanism of M-shape, LinkTouch, introduced in \cite{tsetserukou2014} and further developed in \cite{altamirano2019linkglide}. This mechanism enables sensations of tapping and sliding along the trajectory of the end-effector created by the device, providing multi-contact tactile stimuli on the user's forearm.

To ensure a more natural experience, the device was ergonomically designed to wrap around the forearm. This placement allows users to keep their hands free for other activities while interacting with the device and receiving tactile music sensations. Additionally, the visible nature of the device allows external users to recognize that the user is engaged in an interactive experience.

The system utilizes an array of inverted five-bar linkages to simulate intricate tactile interactions on the user's forearm. As shown in Fig. \ref{fig:device}, the three inverse five-bar linkage mechanisms are arranged parallel to the user's arm, enabling the sensations to be delivered along the forearm.



\begin{figure}[]
    \centering
    \begin{subfigure}[b]{0.45\textwidth}
        \centering
        \includegraphics[width=\textwidth]{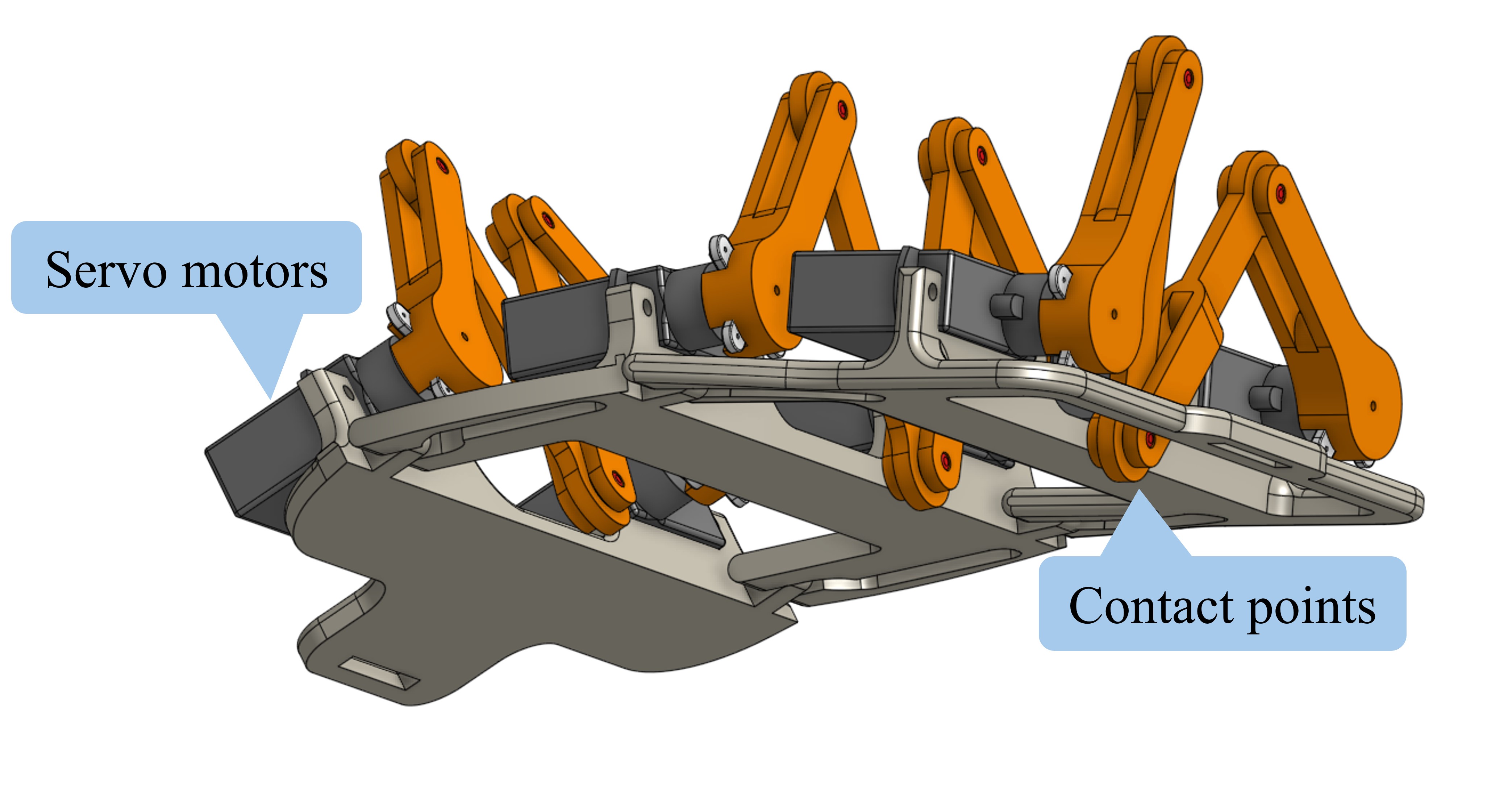}
        \caption{}
        \label{fig:sub1}
    \end{subfigure}
    \hfill
    \begin{subfigure}[b]{0.4\textwidth}
        \centering
        \includegraphics[width=\textwidth]{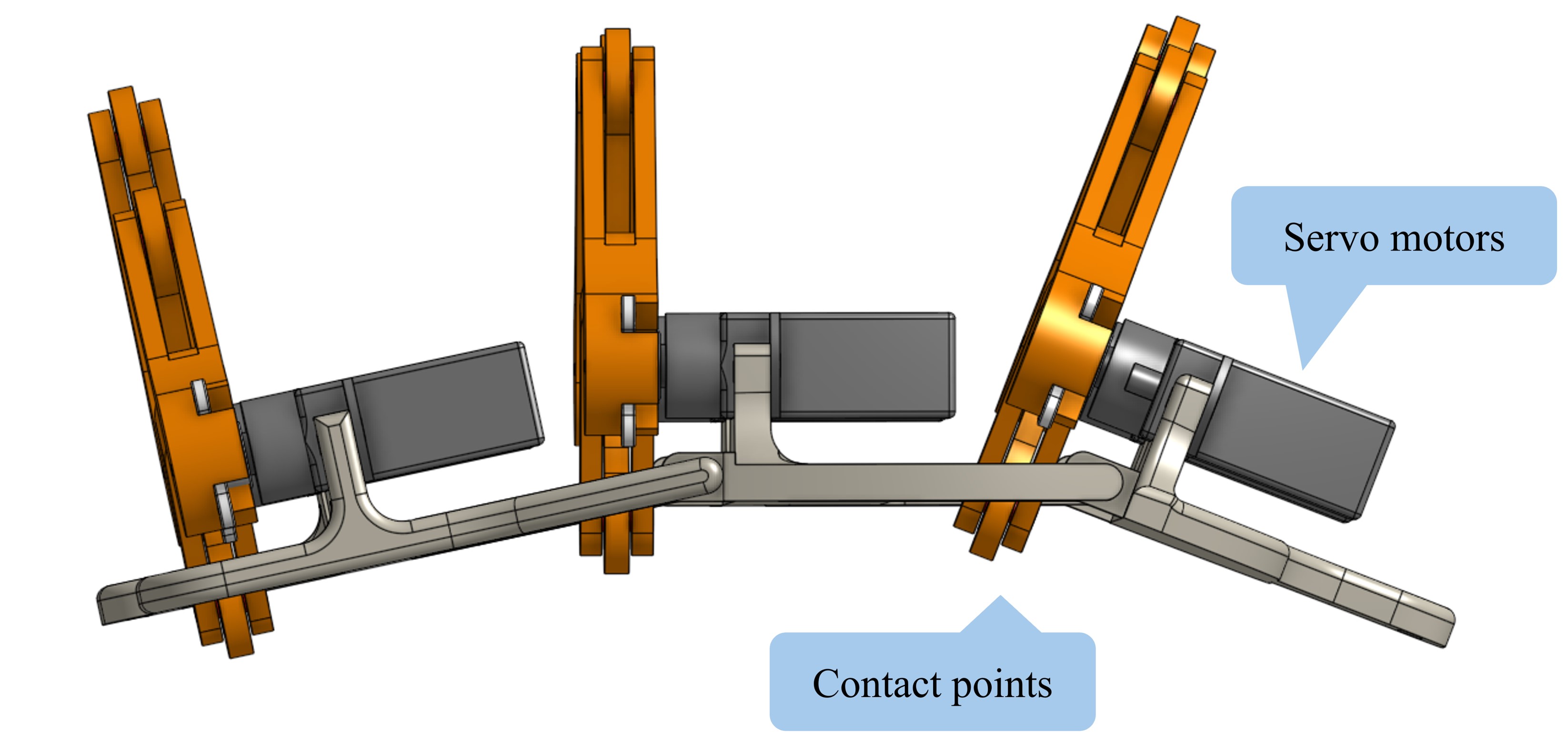}
        \caption{}
        \label{fig:sub2}
    \end{subfigure}
    \begin{subfigure}[b]{0.4\textwidth}
        \centering
        \includegraphics[width=\textwidth]{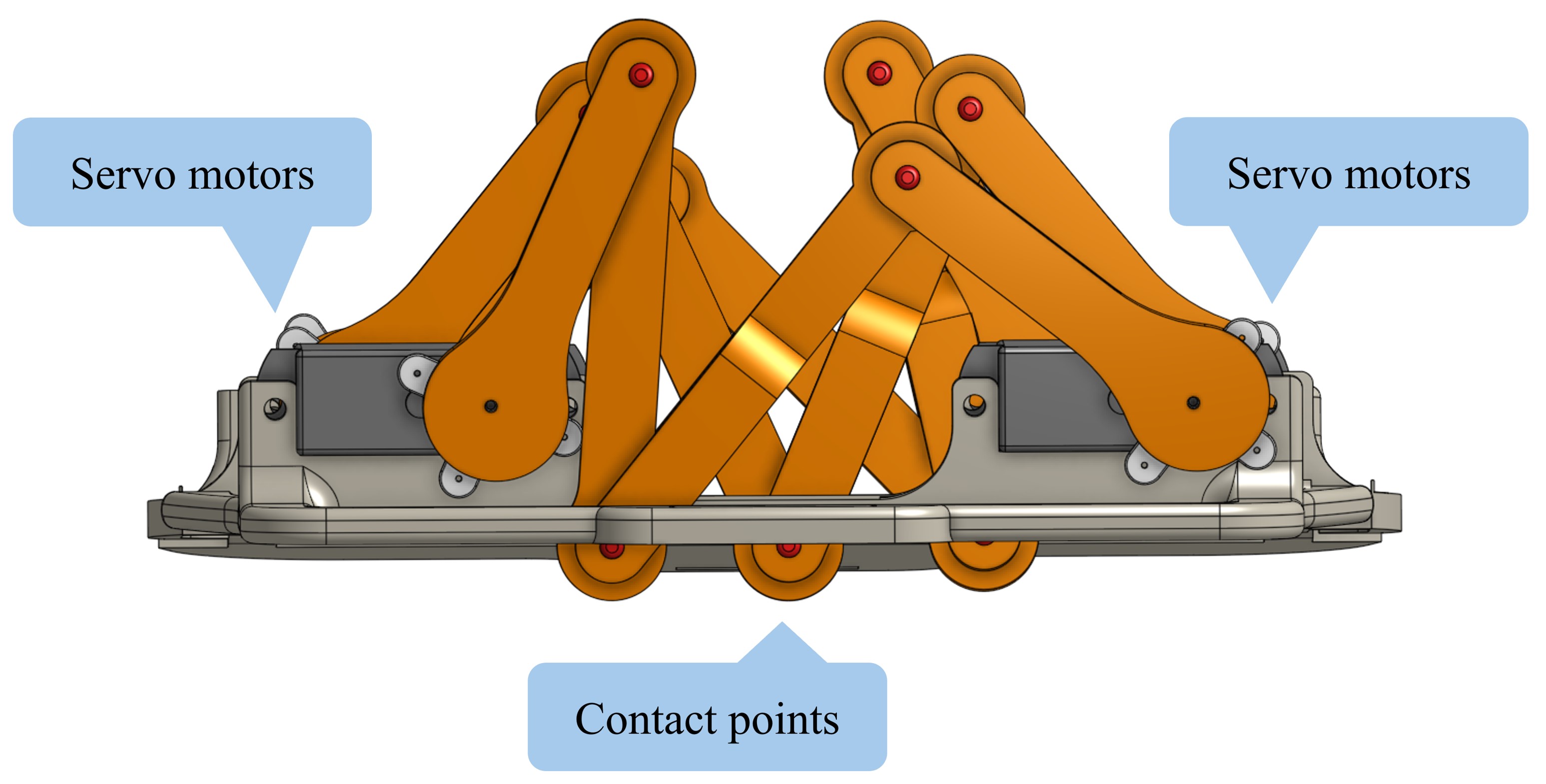}
        \caption{}
        \label{fig:sub3}
    \end{subfigure}
    \caption{3D model of the haptic interface based on the inverted five-bar linkage mechanism. (a) isometric view, (b) lateral view, and (c) front view.}
    \label{fig:inverted}
\end{figure}


\section{EXPERIMENTAL EVALUATION}

We conducted a series of evaluations to determine the human recognition of the bits from the local user while performing the beats from different songs. 

The experiment on tactile perception is centered on the analysis of the human perception of tactile rendering. To study the interference of the noise produced by the servo motors while the wearable device is activated, during the first part of the experiment, the remote users wore the device on the forearm without noise cancellation. During the second part of the experiment, the participants wore headphones with white noise to cancel the noise produced by the device. 

\begin{figure}[t!]
 \centering
 \includegraphics[width=0.48\textwidth]{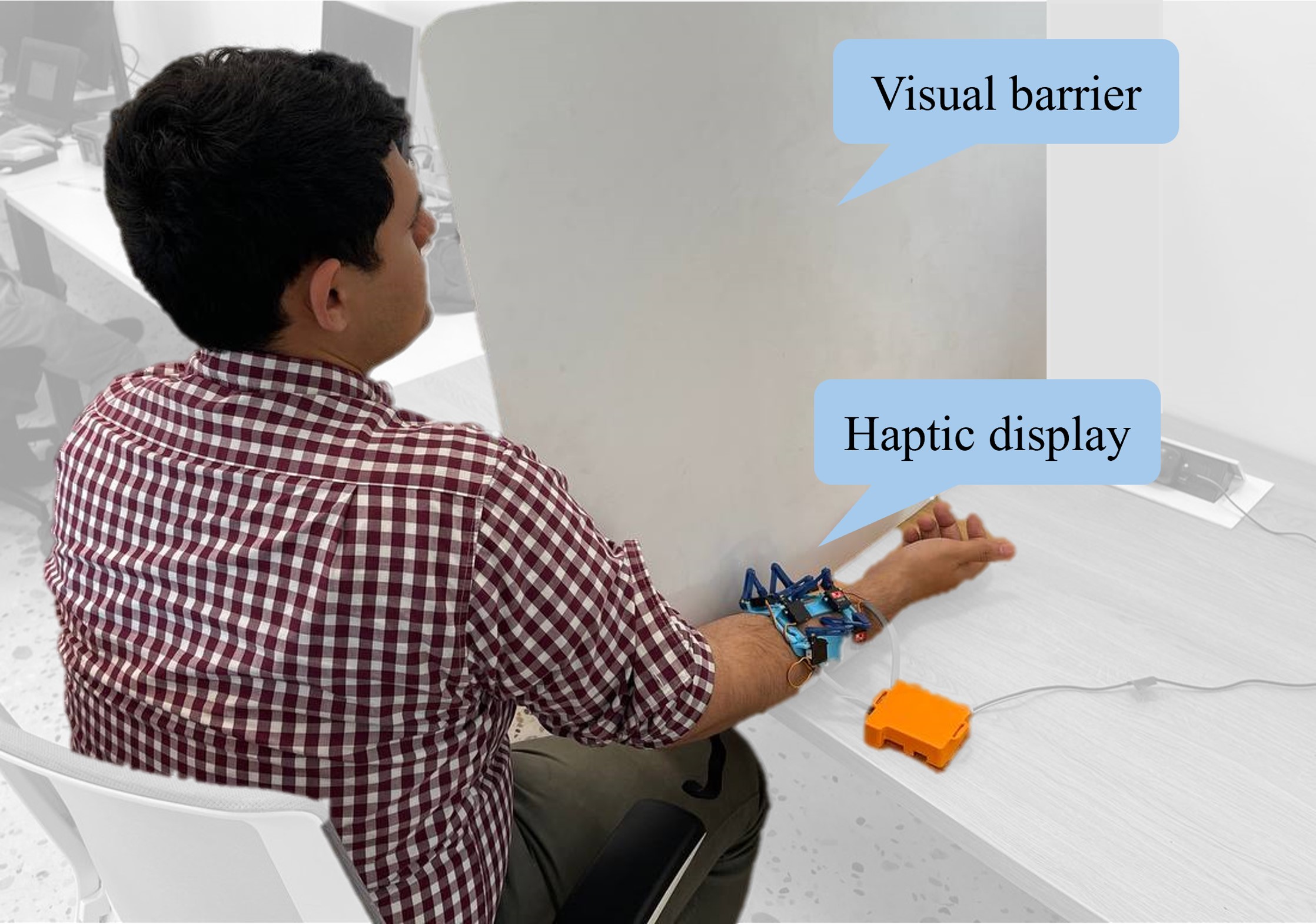}
 \caption{Experimental setup  (Wearable haptic display). The user is wearing the haptic device on the right forearm and telling the patterns perceived.}
 \label{fig:remote}
\end{figure}

\begin{figure}[h!]
 \centering
 \includegraphics[width=0.48\textwidth]{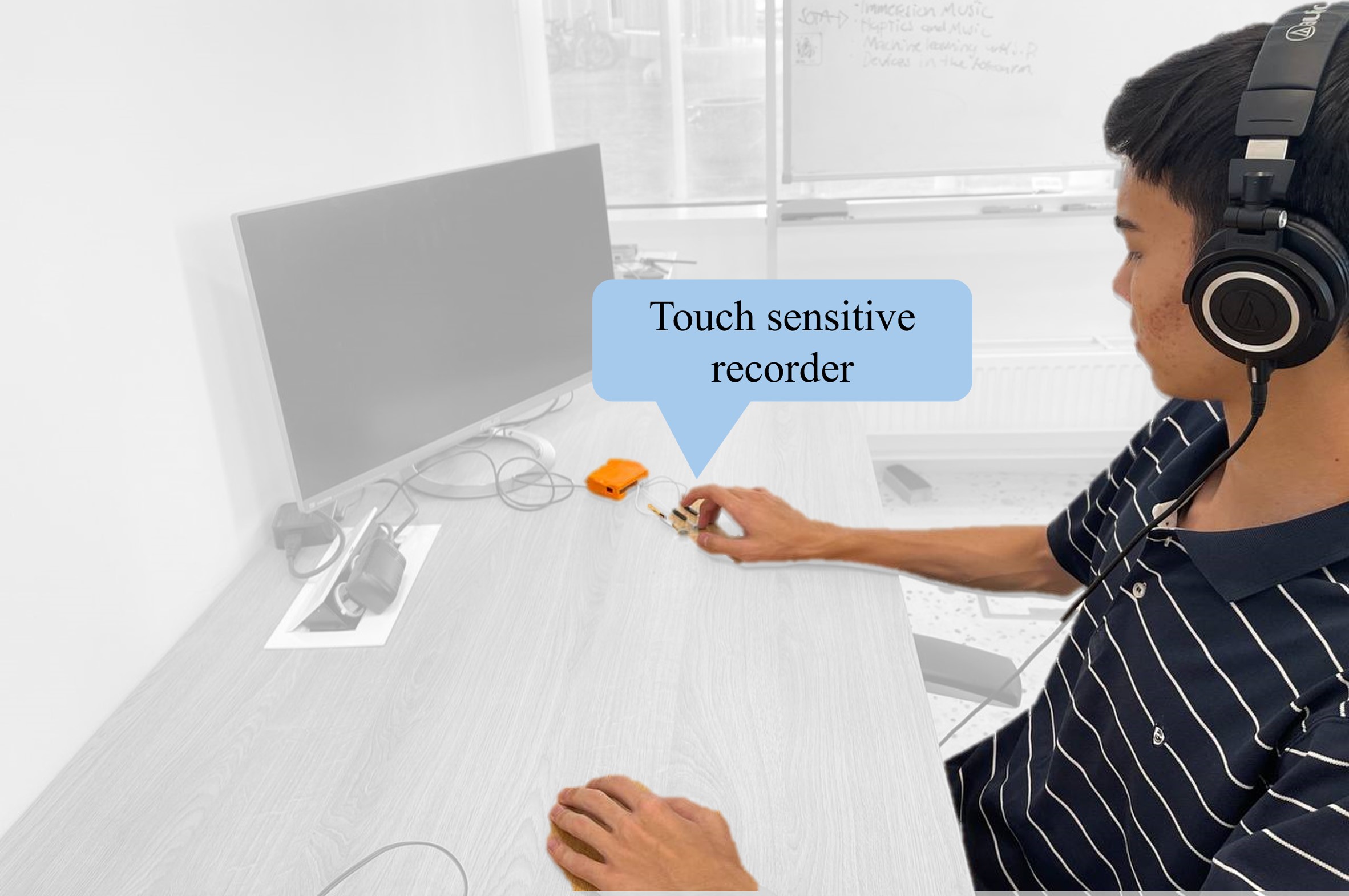}
 \caption{Experimental setup (Touch-sensitive recorder). The user is performing the haptic patterns with their fingers interacting with the three touch sensors located on the table. }
 \label{fig:local}
\end{figure}

Eight right-handed participants (three females) aged 22 to 47 (28.1  ± 8.09) years volunteering completed the evaluation. None of them reported any deficiencies in sensorimotor function.  

Four very known songs were selected to be rendered by the haptic device: (a) Baby Shark, (b) Happy Birth Day, (c) Jingle Bells, and (d) William Tell Overture Final (Horse melody).

Before the study, the experimenter explained the purpose of the multi-contact haptic device to each participant and demonstrated the tactile feedback provided by the device. A demonstration of the melodies was provided. During the experiment, the user was asked to sit in front of a desk and to wear the haptic display on the right forearm as shown in Fig. \ref{fig:remote}. Each melody was presented 3 times blindly in random order, thus, 12 patterns were provided to each participant in each evaluation, once without white noise and once with white noise. Fig. \ref{fig:local} illustrates the Touch-sensitive recorder.

\begin{table}[]
\label{table:confusion_with_noWN}
\caption{Confusion Matrix for Actual and Perceived Melodies Across All Subjects for Recognition without White Noise.}
\centering{
\begin{tabular}{|cc|llll|}
\hline
\multicolumn{2}{|c|}{\cellcolor[HTML]{FFFFFF}}                     & \multicolumn{4}{l|}{\textit{Answers   (Predicted Class)}}                                                                                                                                \\ \cline{3-6} 
\multicolumn{2}{|c|}{\multirow{-2}{*}{\cellcolor[HTML]{FFFFFF}\%}} & \multicolumn{1}{c|}{A}                            & \multicolumn{1}{c|}{B}                            & \multicolumn{1}{c|}{C}                            & \multicolumn{1}{c|}{D}       \\ \hline
\multicolumn{1}{|c|}{}                                       & A   & \multicolumn{1}{l|}{\cellcolor[HTML]{3279C0}0.96} & \multicolumn{1}{l|}{\cellcolor[HTML]{F7FAFD}0.04} & \multicolumn{1}{l|}{\cellcolor[HTML]{FFFFFF}0.00} & \cellcolor[HTML]{FFFFFF}0.00 \\ \cline{2-6} 
\multicolumn{1}{|c|}{}                                       & B   & \multicolumn{1}{l|}{\cellcolor[HTML]{FFFFFF}0.00} & \multicolumn{1}{l|}{\cellcolor[HTML]{2973BD}1.00} & \multicolumn{1}{l|}{\cellcolor[HTML]{FFFFFF}0.00} & \cellcolor[HTML]{FFFFFF}0.00 \\ \cline{2-6} 
\multicolumn{1}{|c|}{}                                       & C   & \multicolumn{1}{l|}{\cellcolor[HTML]{FFFFFF}0.00} & \multicolumn{1}{l|}{\cellcolor[HTML]{F7FAFD}0.04} & \multicolumn{1}{l|}{\cellcolor[HTML]{3279C0}0.96} & \cellcolor[HTML]{FFFFFF}0.00 \\ \cline{2-6} 
\multicolumn{1}{|c|}{\multirow{-4}{*}{\textit{Patterns}}}    & D   & \multicolumn{1}{l|}{\cellcolor[HTML]{FFFFFF}0.00} & \multicolumn{1}{l|}{\cellcolor[HTML]{FFFFFF}0.00} & \multicolumn{1}{l|}{\cellcolor[HTML]{FFFFFF}0.00} & \cellcolor[HTML]{2973BD}1.00 \\ \hline
\end{tabular}}
\end{table}

The results of the human perception evaluation by rendering the melodies are summarized in confusion matrices. Table I corresponds to the answers without white noise, and Table II to the answers with white noise.

\begin{table}[]
\label{table:confusion_with_noWN}
\caption{Confusion Matrix for Actual and Perceived Melodies Across All Subjects for Recognition with White Noise.}
\centering{
\begin{tabular}{|cc|llll|}
\hline
\multicolumn{2}{|c|}{\cellcolor[HTML]{FFFFFF}}                     & \multicolumn{4}{l|}{\textit{Answers   (Predicted Class)}}                                                                                                                                \\ \cline{3-6} 
\multicolumn{2}{|c|}{\multirow{-2}{*}{\cellcolor[HTML]{FFFFFF}\%}} & \multicolumn{1}{c|}{A}                            & \multicolumn{1}{c|}{B}                            & \multicolumn{1}{c|}{C}                            & \multicolumn{1}{c|}{D}       \\ \hline
\multicolumn{1}{|c|}{}                                       & A   & \multicolumn{1}{l|}{\cellcolor[HTML]{3279C0}0.96} & \multicolumn{1}{l|}{\cellcolor[HTML]{FFFFFF}0.00} & \multicolumn{1}{l|}{\cellcolor[HTML]{F7FAFD}0.04} & \cellcolor[HTML]{FFFFFF}0.00 \\ \cline{2-6} 
\multicolumn{1}{|c|}{}                                       & B   & \multicolumn{1}{l|}{\cellcolor[HTML]{FFFFFF}0.00} & \multicolumn{1}{l|}{\cellcolor[HTML]{3B7FC3}0.92} & \multicolumn{1}{l|}{\cellcolor[HTML]{EEF4FA}0.08} & \cellcolor[HTML]{FFFFFF}0.00 \\ \cline{2-6} 
\multicolumn{1}{|c|}{}                                       & C   & \multicolumn{1}{l|}{\cellcolor[HTML]{F7FAFD}0.04} & \multicolumn{1}{l|}{\cellcolor[HTML]{E5EEF7}0.13} & \multicolumn{1}{l|}{\cellcolor[HTML]{4D8BC8}0.83} & \cellcolor[HTML]{FFFFFF}0.00 \\ \cline{2-6} 
\multicolumn{1}{|c|}{\multirow{-4}{*}{\textit{Patterns}}}    & D   & \multicolumn{1}{l|}{\cellcolor[HTML]{FFFFFF}0.00} & \multicolumn{1}{l|}{\cellcolor[HTML]{FFFFFF}0.00} & \multicolumn{1}{l|}{\cellcolor[HTML]{FFFFFF}0.00} & \cellcolor[HTML]{2973BD}1.00 \\ \hline
\end{tabular}}
\end{table}

In order to evaluate the statistical significance of the differences between the perception of the haptic rendering of the melodies (4 songs), we analyzed the results using single factor repeated-measures ANOVA, with a chosen significance level of $\alpha<0.05$. The open-source statistical packages Pingouin and Stats models were used for the statistical analysis.

According to the ANOVA results, there is not a statistically significant difference in the recognition songs without white noise $F(3, 28) = 0.666, p = 0.5795$ neither with white noise $F(3, 28) = 0.9839, p = 0.4144$. The ANOVA showed that the music does not significantly influence the percentage of correct responses separately, with and without noise.

To evaluate the influence of the white noise in the perception of the music by listening to the servo motors moving in the device, we analyzed the results using two-factor repeated-measures ANOVA, with a chosen significance level of $\alpha<0.05$. According to the results, there is no statistically significant difference in the recognition songs without white noise $F(3, 57) = 0.649, p = 0.586$

The overall recognition rate is $98\%$ without white noise and $93\%$ with white noise, which means that the user can distinguish the rhythms of the proposed melodies.

\section{APPLICATIONS}
\begin{itemize}
    \item \textbf{Music Communication through Tactile Feedback:} Haptic technology can transform auditory musical signals into tactile sensations, allowing for innovative forms of musical communication. By translating sound into touch, individuals can perceive rhythm and melody through physical sensations. This application holds particular significance in environments where auditory feedback is limited or inaccessible, as it provides a new modality for engaging with music. Haptic feedback systems can bridge the gap between auditory and tactile perception, offering an alternative way to experience and interact with music.
    \item \textbf{Remote Music Collaboration:} Haptic feedback facilitates real-time remote musical collaboration by transmitting tactile cues between distant performers. Musicians can feel the rhythm and dynamics of their collaborators’ performances, enabling more synchronized and expressive interactions despite physical separation. This technology can be especially valuable in a globalized and digitally connected world, where remote collaborations are increasingly common. By incorporating haptic feedback, remote musical interactions can achieve a level of immediacy and connectedness that closely mirrors in-person performances.
    \item \textbf{Accessibility for the Hearing Impaired:} This technology offers transformative potential in making musical experiences accessible to individuals with hearing impairments. By converting sound into vibrations and tactile stimuli, haptic devices allow users to perceive music through their sense of touch. This approach enables those who are hard of hearing to engage with musical elements such as rhythm, intensity, and emotional expression. The application of haptic feedback in this context not only enhances inclusivity but also expands the sensory possibilities for experiencing music.
    \item \textbf{Immersive and Multisensory Musical Experiences:} This technology can enhance the immersive quality of musical experiences by adding a tactile component to auditory stimuli. In virtual reality (VR) environments, live performances, or interactive installations, haptic feedback can simulate the physical sensations associated with music, such as the vibrations of instruments or the pulsations of bass frequencies. This multisensory integration deepens the user's engagement with music, creating a more holistic and embodied experience. By combining auditory and tactile inputs, haptic technology enriches the overall sensory impact of music.
   \item \textbf {Tactile Music Perception and Education:} Haptic feedback systems can be employed to convey musical information through touch, allowing users to recognize and differentiate between various musical elements such as rhythm, tempo, and melody. This tactile approach to music perception has significant implications for both education and entertainment. In educational settings, haptic feedback can be used as a tool to teach and reinforce musical concepts, particularly for students who may benefit from multisensory learning strategies. In entertainment, it adds a new dimension to how music is experienced, making it more engaging and interactive.
\end{itemize}
\section{CONCLUSION AND FUTURE WORK}

This paper explored a novel system that translates auditory musical signals into haptic feedback, enabling users to perceive music through tactile sensations. Through a series of evaluations, our study demonstrated that participants could effectively recognize and differentiate between familiar melodies using a wearable haptic device, even when auditory feedback was absent. With experimental results showing high recognition rates of $98\%$ without noise and $93\%$ with noise, the system's capabilities in conveying complex musical information are substantiated.

The findings have significant implications for various applications, including enhancing music accessibility for individuals with hearing impairments, facilitating remote musical collaboration, and enriching multisensory experiences in entertainment and education. By bridging the gap between auditory and tactile perception, this research contributes to the growing field of musical haptics and offers new possibilities for how we engage with music.

Future work will focus on refining the system’s design for broader usability and exploring its application in diverse musical genres and performance contexts. Additionally, extending the research to more complex musical structures and longer compositions will help to further validate the system’s effectiveness. Efforts will be directed towards improving the precision and range of the haptic feedback to enhance the fidelity of musical reproduction. Additionally, we are planning to develop and implement the MusingerGPT system, which will be based on a Large Language Model to generate distinctive tactile patterns for musical compositions. Lastly, more extensive user studies involving diverse demographics will provide deeper insights into the usability and overall impact of the system in real-world settings.





\section*{Acknowledgements} 
Research reported in this publication was financially supported by the RSF grant No. 24-41-02039.







\begin{thebibliography}{xx}

\bibitem{Papetti2018}
S. Papetti and C. Saitis,\textit{ Musical Haptics: Introduction}, pp. 1–7. Cham: Springer International Publishing, 2018.

\bibitem{Berdahl2018}
E. Berdahl, A. Pfalz, M. Blandino, and S. D. Beck,\textit{ Force-Feedback Instruments for the Laptop Orchestra of Louisiana}, pp. 171–191. Cham: Springer International Publishing, 2018.

\bibitem{Altukhaim2024}
S. Altukhaim, D. George, K. Nagaratnam, T. Kondo, and Y. Hayashi, “Enhancement of sense of ownership using virtual and haptic feedback,”\textit{ Scientific Reports}, vol. 14, p. 5140, Mar 2024.

\bibitem{Tivadar2022}
R. I. Tivadar, R. C. Arnold, N. Turoman, J.-F. Knebel, and M. M. Murray, “Digital haptics improve speed of visual search performance in a dual-task setting,” \textit{Scientific Reports}, vol. 12, p. 9728, Jun 2022.

\bibitem{turchet2021touching}
L. Turchet, T. West, and M. M. Wanderley, “Touching the audience: musical haptic wearables for augmented and participatory live music performances,” \textit{Personal and Ubiquitous Computing}, vol. 25, pp. 749–
769, 2021.

\bibitem{grosshauser2009augmented}
T. Grosshauser and T. Hermann, “Augmented haptics – an interactive feedback system for musicians,” \textit{in Proc. Int. Conf. Haptic and Audio Interaction Design: HAID}, pp. 100–108, 2009.

\bibitem{frid2020haptic}
E. Frid and H. Lindetorp, “Haptic music: exploring whole-body vibrations and tactile sound for a multisensory music installation,” \textit{in Proc. Sound and Music Computing Conference}, pp. 68–75, 2020.

\bibitem{armitage2015configuring}
J. Armitage and K. Ng, “Configuring a haptic interface for music performance,” \textit{in Electronic Visualisation and the Arts (EVA)}, BCS Learning \& Development, 2015.

\bibitem{mazzoni2016mood}
A. Mazzoni and N. Bryan-Kinns, “Mood glove: A haptic wearable prototype system to enhance mood music in film,” \textit{Entertainment Computing}, vol. 17, pp. 9–17, 2016.

\bibitem{muller2010reflective}
A. M{\"u}ller, F. Hemmert, G. Wintergerst, and R.  Jagodzinski, “Reflective haptics : Resistive force feedback for musical performances with stylus-controlled instruments,” \textit{in Proc. Int. Conf. on New Interfaces for Musical Expression}, (Sydney, Australia), pp. 477–478, 2010.

\bibitem{kirkegaard2020torquetuner}
M. Kirkegaard, M. Bredholt, C. Frisson, and M. M. Wanderley, “Torquetuner: A self contained module for designing rotary haptic force feedback for digital musical instruments,”  \textit{in in Proc. Int. Conf. on New Interfaces for Musical Expression}, 2020

\bibitem{heredia2019recyglide}
J. Heredia, J. Tirado, V. Panov, M. Altamirano Cabrera, K. Youcef-Toumi, and D. Tsetserukou, “Recyglide: A forearm-worn multi-modal haptic display aimed to improve user vr immersion submission,” \textit{in Proc. of the 25th ACM Symposium on Virtual Reality Software and Technology}, pp. 1–2, 2019.

\bibitem{10.1145/3355355.3361896}
D. Trinitatova and D. Tsetserukou, “Touchvr: a wearable haptic interface for vr aimed at delivering multi-modal stimuli at the user’s palm,” \textit{in SIGGRAPH Asia XR}, p. 42–43, 2019.

\bibitem{10.1007/978-3-642-14064-8_49}
D. Tsetserukou, “Haptihug: A novel haptic display for communication of hug over a distance,” \textit{in Haptics: Generating and Perceiving Tangible Sensations}, pp. 340–347, 2010.

\bibitem{6775473}
D. Tsetserukou, S. Hosokawa, and K. Terashima, “Linktouch: A wearable haptic device with five-bar linkage mechanism for presentation of two-dof force feedback at the fingerpad,” \textit{in In Proc. 2014 IEEE Haptics Symposium (HAPTICS)}, pp. 307–312, 2014.

\bibitem{5349516}
D. Tsetserukou, A. Neviarouskaya, H. Prendinger, N. Kawakami, and S. Tachi, “Affective haptics in emotional communication,” \textit{in 3rd Int. Conf. on Affective Computing and Intelligent Interaction and Workshops}, pp. 1–6, 2009.

\bibitem{alabbas2024movetouch}
A. Alabbas, M. A. Cabrera, M. Sayed, O. Alyounes, Q. Liu, and D. Tsetserukou, “Movetouch: Robotic motion capturing system with wearable
tactile display to achieve safe hri,”\textit{ in Int. Conf. EuroHaptics}, pp. 1-6, 2024

\bibitem{altamirano2019linkglide}
M. Altamirano Cabrera and D. Tsetserukou, “Linkglide: a wearable haptic display with inverted five-bar linkages for delivering multi-contact and multi-modal tactile stimuli,” \textit{in Haptic Interaction: Perception,
Devices and Algorithms 3}, pp. 149–154, Springer, 2019.


\end{thebibliography}
\end{document}